\documentclass[12pt]{article}

\usepackage{graphicx}
\usepackage{amsmath,amssymb,bm}
\usepackage{gensymb}
\usepackage{nicefrac} 
\usepackage{cite}
\usepackage[normalem]{ulem}
\usepackage{color}
\usepackage{slashed}
\usepackage{mathrsfs}
\usepackage{url}
\usepackage{xspace} 
\usepackage[bookmarks, breaklinks, colorlinks,urlcolor=black, citecolor=red, linkcolor=blue]{hyperref}
\usepackage{braket}
\usepackage[dvipsnames]{xcolor}

\usepackage{tikz}
\usetikzlibrary{arrows.meta}
\makeatletter
\newlength\lrvec@height
\newlength\lrvec@width
\newif\iflrvec@same@height
\def\lrvec{\@ifstar\slrvec@\lrvec@}
\newcommand{\slrvec@}[2][.4ex]{
  \lrvec@same@heighttrue
  \mathpalette\lrvec@@{{#1}{#2}}
}
\newcommand{\lrvec@}[2][.4ex]{
  \lrvec@same@heightfalse
  \mathpalette\lrvec@@{{#1}{#2}}
}
\def\lrvec@@#1#2{\lrvec@@@#1#2}
\def\lrvec@@@#1#2#3{%
  \iflrvec@same@height
    \settoheight{\lrvec@height}{$\m@th#1 \mathbf{T}#3$}
  \else
    \settoheight{\lrvec@height}{$\m@th#1#3$}
  \fi
  \settowidth{\lrvec@width}{$\m@th#1#3$}
  \kern.08em
  \raisebox{#2}{\raisebox{\lrvec@height}{\rlap{%
    \kern-.05em
    \begin{tikzpicture}[<-> /.tip={To[width=.4em, length=.2em]}]
      \draw [<->] (-.05em,0)--(\lrvec@width+.05em,0);
    \end{tikzpicture}%
  }}}%
  #3
  \kern.08em
}
\makeatother

\def\beq{\begin{equation}}
\def\eeq{\end{equation}}
\def\bea{\begin{eqnarray}}
\def\eea{\end{eqnarray}}
\def\nn{\nonumber}

\def\roughly#1{\mathrel{\raise.3ex\hbox
{$#1$\kern-.75em\lower1ex\hbox{$\sim$}}}}
\def\lsim{\roughly<}
\def\gsim{\roughly>}

\def\sla#1{\raise.15ex\hbox{$/$}\kern-.57em #1}
\def\bsmumu{b \to s \mu^+ \mu^-}
\def\bsnunubar{b \to s \nu {\bar\nu}}
\def\bctaunu{b \to c \tau^- {\bar\nu}}
\def \cB{{\cal B}}

\begin{document}

\begin{flushright}
UdeM-GPP-TH-22-294 \\
\end{flushright}

\begin{center}
\bigskip
{\Large \bf \boldmath The $B$ Anomalies, the $U_1$ Leptoquark \\ and Dark Matter} 
\\
\bigskip
\bigskip
{\large
Genevi\`eve B\'elanger $^{a,}$\footnote{belanger@lapth.cnrs.fr},
Jacky Kumar $^{b,c,}$\footnote{jacky.kumar@tum.de},
David London $^{b,}$\footnote{london@lps.umontreal.ca}
and Alexander Pukhov $^{d,}$\footnote{alexander.pukhov@gmail.com}
}
\end{center}

\begin{flushleft}
~~~~~~~~~~~$a$: {\it LAPTh, CNRS, USMB, 9 Chemin de Bellevue, 74940 Annecy, France}\\
~~~~~~~~~~~$b$: {\it Physique des Particules, Universit\'e de Montr\'eal,}\\
~~~~~~~~~~~~~~~{\it 1375 Avenue Th\'er\`ese-Lavoie-Roux, Montr\'eal, QC, Canada  H2V 0B3}\\
~~~~~~~~~~~$c$: {\it Institute for Advanced Study, Technical University Munich,}\\
~~~~~~~~~~~~~~~{\it Lichtenbergstr.\ 2a, D-85747 Garching, Germany}\\
~~~~~~~~~~~$d$: {\it Skobeltsyn Institute of Nuclear Physics,}\\
~~~~~~~~~~~~~~~{\it Moscow State University, Moscow 119992, Russia } \\
\end{flushleft}

\begin{center}
\bigskip (\today)
\vskip0.5cm {\Large Abstract\\} 
\vskip3truemm
\parbox[t]{\textwidth}{The present-day $B$-anomalies involving
  $\bsmumu$ or $\bctaunu$ transitions can all be explained with the
  addition of a vector $U_1$ leptoquark with a mass of $M_{U_1} \gsim
  1.8$ TeV. In the scalar singlet dark matter model (SSDMM), the DM is
  a scalar $S$ that couples to the Higgs via $\lambda_{hS} \, S^2
  |H|^2$. We update the fit to the data and find that the SSDMM is now
  viable only for $M_S \gsim 1.6$ TeV. In this paper, we assume that
  the DM also couples to the $U_1$ via $\lambda_{U_1 S} \, S^2 \, U_{1
    \mu}^{\dagger} U^{\mu}_1$. In addition to leading to DM
  annihilation via $S S \to U_1 {\bar U}_1$, this coupling generates
  $SSgg$ and $SS\gamma\gamma$ couplings at one loop. Although naively
  divergent, these loop diagrams can be calculated under the
  assumption that the $U_1$ is a gauge boson of a group broken at the
  TeV scale. With this DM-$U_1$ coupling term, there are additional
  contributions to the various DM observables (relic density, direct
  and indirect detection). We find that the constraints on the SSDMM
  are relaxed for both heavy DM ($M_S \gsim M_{U_1}$) and light DM
  ($M_S < M_{U_1}$).}
\end{center}

\thispagestyle{empty}
\newpage
\setcounter{page}{1}
\baselineskip=14pt

\section{Introduction}

At the present time, there are discrepancies with the predictions of
the standard model (SM) in measurements of a variety of observables in
$B$ decays (for a review, see Ref.~\cite{London:2021lfn}). First,
there are observables involving the neutral-current transition
$\bsmumu$. These include the branching ratios and angular
distributions of $B \to K^* \mu^+ \mu^-$ and $B_s \to \phi \mu^+
\mu^-$, ${\cal B}(B_s\to\mu^+\mu^-)$, and the
lepton-flavour-universality-violating (LFUV) ratios $R_{K^{(*)}}
\equiv \cB(\bar{B} \to K^{(*)} \mu^+ \mu^-) / \cB(\bar{B} \to K^{(*)}
e^+ e^-)$.  When all the data are combined in a global fit with close
to 250 $\bsmumu$ observables \cite{Alguero:2021anc}, it is found that
the overall disagreement with the SM is at the level of $2.5\sigma$. A
variety of new-physics (NP) models have been proposed to explain this,
the simplest involving the addition of a $Z'$ boson or a (scalar or
vector) leptoquark (LQ) \cite{London:2021lfn}.

There are also anomalies in observables involving the charged-current
decay $\bctaunu$ \cite{London:2021lfn}, in particular the LFUV ratios
$R_{D^{(*)}} \equiv \cB(\bar{B} \to D^{(*)} \tau^{-} {\bar\nu}_\tau) /
\cB(\bar{B} \to D^{(*)} \ell^{-} {\bar\nu}_\ell)$ $(\ell = e,\mu)$ and
$R_{J/\psi} \equiv \cB(B_c^+ \to J/\psi\tau^+\nu_\tau) / \cB(B_c^+ \to
J/\psi\mu^+\nu_\mu)$.  Here the discrepancy is somewhat larger, at the
level of $\sim 3.3\sigma$ \cite{Blanke:2019qrx}. Once again, various
NP explanations have been proposed, with most involving a $W'$ boson,
a LQ, or a charged Higgs boson \cite{London:2021lfn}.

In Ref.~\cite{Bhattacharya:2014wla}, it was pointed out that it is
possible to construct a NP model that explains both kinds of $B$
anomalies. In such a model, the NP contributions to neutral-current
and charged-current processes would have to be related, which can
occur if the NP couplings are left-handed. Possible NP particles
include a vector-boson triplet $(W',Z')$, as well as several different
types of LQs.

These models were analyzed in Refs.~\cite{Bhattacharya:2016mcc,
  Buttazzo:2017ixm, Kumar:2018kmr}. Although they can in principle
explain the $B$ anomalies, all of these models also contribute to
other processes, whose measurements place constraints on the models.
These include $B_S^0$-${\bar B}_s^0$ mixing, $b \to s {\bar\nu}\nu$
and $\tau\to\mu\phi$ decays, etc.  When all such constraints are taken
into account, as well as those from direct searches at the LHC, only
one model survives. It uses a vector $SU(2)_L$-singlet leptoquark (LQ)
of charge $-2/3$, known as $U_1$. Thus, all $B$ anomalies can be
explained with the addition of the $U_1$ LQ\footnote{Another (somewhat
  less-elegant) possibility is to add the two scalar LQs $S_1$ and
  $S_3$ \cite{Crivellin:2017zlb}.  The $SU(2)_L$-triplet $S_3$
  contributes to both $\bsmumu$ and $\bctaunu$, while the
  $SU(2)_L$-singlet $S_1$ contributes only to $\bsmumu$.  By carefully
  choosing the couplings of $S_1$ and $S_3$, both anomalies can be
  explained, while evading the constraints from $\bsnunubar$.}.
  
Another observation that requires physics beyond the SM is dark matter
(DM).  Although DM is the most prevalent form of matter in the
universe, we know very little about its nature. It must be neutral (or
very nearly so), but we have no knowledge about its mass or spin. It
was present in the early universe, so its presence today indicates
that it must be stable.  This is most likely due to a symmetry,
usually taken to be $Z_2$, under which the DM and SM particles have
opposite charges. The relic density of DM is determined precisely from
the observations of the Cosmic Microwave Background
\cite{Planck:2018vyg}.  In the standard freeze-out mechanism, the rate
of DM annihilation into SM particles determines the DM relic
density. For these annihilations to take place, there must be a
mediator that couples to the DM and to some SM particles. The exchange
of this mediator will also produce DM elastic scattering on nuclei,
allowing for the direct detection of DM, and DM annihilation in the
galaxy, leading to indirect detection of DM in cosmic rays.

One of the simplest mediator scenarios is the scalar singlet dark
matter model (SSDMM) \cite{Silveira:1985rk, McDonald:1993ex,
  Burgess:2000yq}.  Here the DM is assumed to be a scalar that couples
to the SM Higgs boson through an $S^2 |H|^2$ term.  This Higgs portal
leads to DM annihilation in the early universe via $S S \to h \to {\rm
  SM~particles}$ and to the scattering with nucleons, $S N \to S N$,
via Higgs exchange, which contributes to direct detection. A global
fit of the SSDMM was performed in 2017 \cite{GAMBIT:2017gge}, and it
was found that the various constraints, particularly the DM relic
density and the upper limit on direct detection, have the effect that
the SSDMM is strongly disfavoured for $M_S \lsim 500$-700 GeV.

Because the $B$ anomalies and DM both offer strong hints of NP, it is
quite natural to consider links between them. Indeed, a number of
papers have proposed models in which the $Z'$ used to explain the
$\bsmumu$ anomalies also acts as the mediator for DM annihilation
\cite{Sierra:2015fma, Belanger:2015nma, Celis:2016ayl,
  Altmannshofer:2016jzy, Ko:2017quv, Ko:2017yrd, Cline:2017lvv,
  Falkowski:2018dsl, Arcadi:2018tly, Hutauruk:2019crc, Biswas:2019twf,
  Han:2019diw, Borah:2020swo}. And a few papers have combined DM with
LQs \cite{Cline:2017aed, Guadagnoli:2020tlx, Baker:2021llj,
  Choi:2018stw, Belanger:2021smw}. In one of these, the two scalar LQs
$S_1$ and $S_3$ provide a portal for the scalar DM particle $S$ via an
$S^2 |S_{LQ}|^2$ coupling ($S_{LQ} = S_1$, $S_3$) \cite{Choi:2018stw}.

In the present paper, we apply this LQ portal idea to the $U_1$ LQ.
For scalar DM $S$, we assume a coupling of the form $S^2 \, U_{1
  \mu}^{\dagger} U^{\mu}_1$. Such a coupling produces the tree-level
annihilation process $S S \to U_1 {\bar U}_1$, which contributes
to the relic density. At one loop, one can also generate an effective
$SSgg$ coupling. This leads to $SS \to gg$ annihilation (relic
density) and contributes to $S N \to S N$ (direct detection). The
problem is that, because the $U_1$ LQ is a vector boson, such loop
diagrams generally diverge. Fortunately, there is a way around this.
The $U_1$ is generally considered to be a gauge boson of an extended
gauge group that is broken at the TeV scale. In this case, the $SSgg$
coupling can be computed by analogy to the SM $hgg$ or $h\gamma\gamma$
couplings. An $SS\gamma\gamma$ coupling, which contributes to indirect
detection signals, is also induced, and can be calculated similarly.

In this paper, we examine the consequences of an $S^2 \, U_{1
  \mu}^{\dagger} U^{\mu}_1$ coupling for the various DM observables --
relic density, direct detection and indirect detection -- for
different DM masses. We treat separately the cases of heavy DM ($M_S
\gsim M_{U_1} = 1.8$ TeV), for which the annihilation process $S S \to
U_1 {\bar U}_1$ is possible, and light DM ($M_S < M_{U_1} = 1.8$ TeV),
for which it is not. In particular, we investigate whether the
addition of the $U_1$ LQ can improve the prospects for the SSDMM. As
we will see, it does do this, for both heavy and light DM.

In Sec.~2, we present the setup. We introduce the effective
Hamiltonian that contains both $U_1$ and $S$, compute the one-loop
coupling term $C_{5}^g S^2 G_{\mu \nu}^a G^{a \mu \nu}$ that yields an
effective $SSgg$ coupling, and describe in more detail the SSDMM.  The
effects on the DM observables are examined in Sec.~3, with heavy and
light DM studied separately.  In the study of light DM, we also update
the constraints on the SSDMM. We find that, in both cases, the
addition of the $U_1$ LQ can indeed make the constraints on the SSDMM
less stringent, thus improving its outlook. We analyze the prospects
for probing this model via DM indirect detection in the photon
channels with the future Cherenkov Telescope Array detector. We
conclude in Sec.~4.

\section{Setup}

The $U_1$ LQ is a vector boson. If one simply adds it to the SM, but
makes no additions to the Higgs sector, the new theory is not
renormalizable. As a consequence, loop diagrams with internal $U_1$
LQs, which can lead to potentially important effects, are generally
divergent (some exceptions can be found in
Ref.~\cite{Crivellin:2018yvo}). To deal with this problem, models
have been constructed with an extended gauge group that contains the
$U_1$ LQ as a gauge boson and is broken at the TeV scale. There are
two types of such UV-completion models: (i) those based on variations
of Pati-Salam models, in which $SU(4)_{PS}$ unifies $SU(3)_C$ and a
$U(1)$ under which both quarks and leptons are charged
\cite{Assad:2017iib, DiLuzio:2017vat, Calibbi:2017qbu,
  Bordone:2017bld, Barbieri:2017tuq, Blanke:2018sro, Aydemir:2018cbb,
  Heeck:2018ntp, Balaji:2018zna, Fornal:2018dqn, Balaji:2019kwe,
  Iguro:2021kdw}, and (ii) those that use the ``4321'' gauge group,
$SU(4) \times SU(3)' \times SU(2)_L \times U(1)_X$
\cite{Guadagnoli:2020tlx, Greljo:2018tuh, Cornella:2019hct}. In our
analysis, we assume that the $U_1$ LQ is a gauge boson of a larger
group, but make no assumptions about the group structure.

\subsection{Effective Hamiltonian}

We take $\Lambda_{NP}$ to be the scale at which this larger gauge
group is broken. Below $\Lambda_{NP}$, but above the
electroweak-breaking scale, the $U_1$ and $S$ are added to the
particle spectrum. At leading order, the effective Lagrangian includes
all operators up to mass dimension four that respect the $SU(3)_C
\times SU(2)_L \times U(1)_Y$ SM gauge symmetry. The full Lagrangian
is
\beq
{\cal L} = {\cal L}_{SM} + {\cal L}_S + {\cal L}_{U_1} + {\cal L}_Y + {\cal L}_{SU_1H} ~,
\eeq
where 
\bea
\label{L_S}
{\cal L}_S &=& \frac12 \partial_\mu S \partial^\mu S - \frac12 \mu_S^2 S^2 - \frac14 \lambda_S S^4 ~, \\
\label{U1U1terms}
    {\cal L}_{U_1} &=& -~\frac{1}{2} (D_\mu U_{1\nu} - D_\nu U_{1\mu}) ^\dagger(D^\mu U^{\nu}_1 - D^\nu U^{\mu}_1)
    - M_{U_1}^2 \, U_{1\mu}^\dagger U_1^\mu \nn\\
&& \hskip2truecm  -~i g_{U_1U_1g} \, U_1^{\dagger\mu}  T^a U_1^\nu  G_{\mu \nu}^a
- i g_{U_1U_1B} \, \frac{2}{3} U_1^{\dagger\mu}   U_1^\nu  B_{\mu \nu} ~, \\
\label{SSU1U1coup}
{\cal L}_Y &=& (g_L^{ij} ~ \bar Q_i \gamma_\mu P_L L_j  + g_{R}^{ij} ~ \bar d_i \gamma_\mu P_R e_j ) ~U_1^{\mu}
+ h.c.,  \\
\label{portals}
{\cal L}_{SU_1H} &=& 
- \frac12 \lambda_{U_1 S} \, S^2 \, U_{1 \mu}^{\dagger} U^{\mu}_1
- \frac14 \lambda_{hS} S^2 |H|^2
- \frac12 \lambda_{U_1 h} \, U_{1 \mu}^{\dagger} U^{\mu}_1 \, |H|^2 ~.
\eea
In ${\cal L}_Y$, $i$ and $j$ are generation indices, and
\begin{equation}
  Q_L =   \begin{pmatrix} V^\dagger u_L  \\ d_L  \end{pmatrix} ~~,~~~~
  L_L =   \begin{pmatrix} \nu_L \\ e_L  \end{pmatrix} ~,
\end{equation}
where $V$ is the Cabibbo-Kobayashi-Maskawa (CKM) matrix.

The $U_1$ portal term $S^2 \, U_{1 \mu}^{\dagger} U^{\mu}_1$
contributes at tree level to the relic density via $S S \to U_1 {\bar
  U}_1$. This process is suppressed when the $U_1$ cannot be produced
on-shell, so that the cross section depends strongly on the mass of
the $U_1$. In Ref.~\cite{Angelescu:2021lln}, using the latest LHC data
and assuming that the LQ couples mainly to the third generation, it
was determined that direct searches at the LHC place a lower limit of
$M_{U_1} = 1.8$ TeV; this is the value we use in our analysis.

\subsection{One-loop DM couplings}

\begin{figure}[!htbp]
\begin{center}
\includegraphics[width=0.40\textwidth]{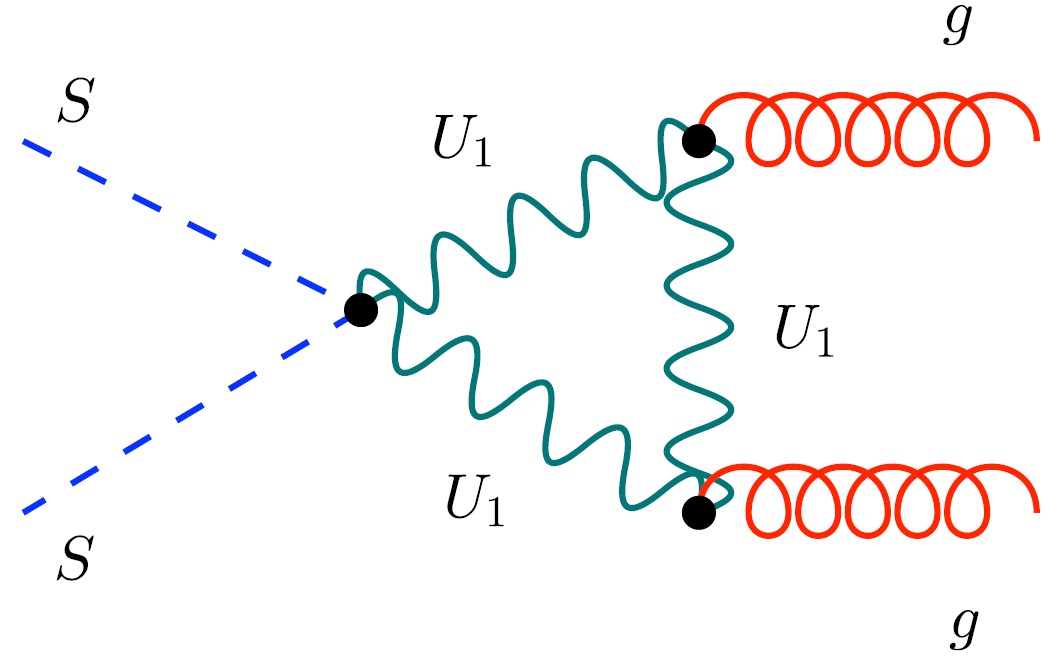}
~~~~~~~
\includegraphics[width=0.40\textwidth]{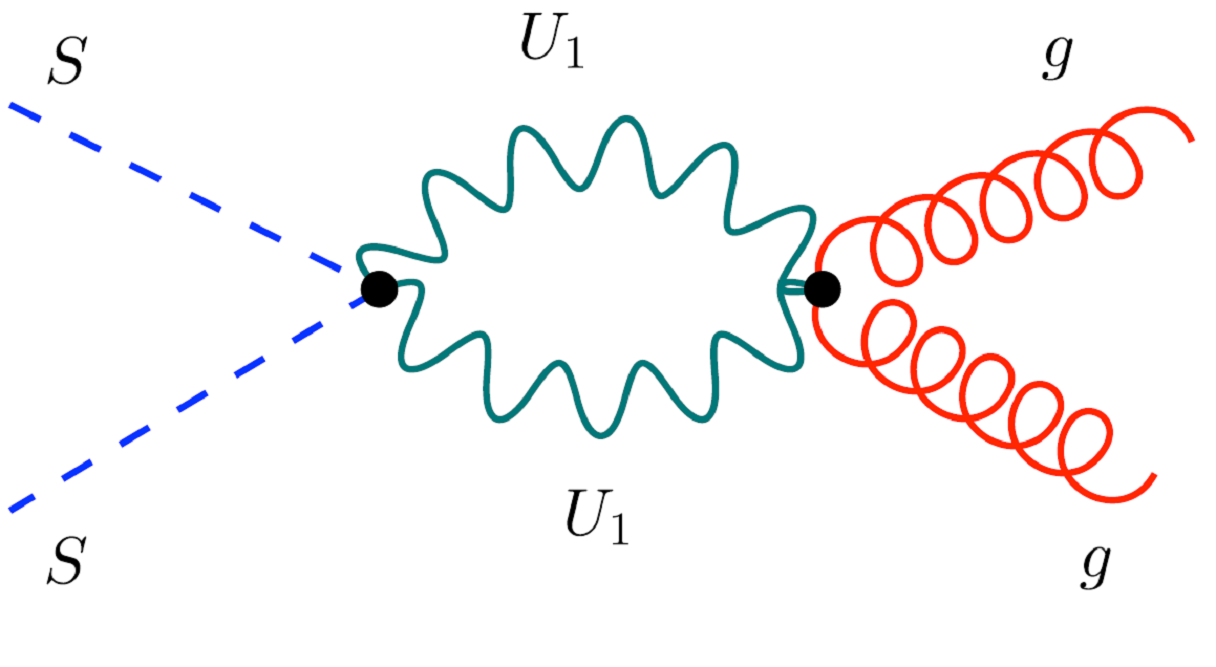}
\end{center}
\vskip-5truemm
\caption{\small Diagrams generating the $SSgg$ coupling.}
\label{SSggcoup}
\end{figure}

Direct detection experiments search for the elastic scattering process
$S N \to S N$. This can be produced using the $U_1$ portal, but only
at the one-loop level. For example, in Ref.~\cite{Godbole:2015gma}, it
is pointed out that, in some models, a coupling of $SS$ to two gluons
can be generated, and that this can mediate the $S N \to S N$
scattering process. Since the $U_1$ LQ is coloured, an $SSgg$ coupling
can be generated at one loop through the diagrams in
Fig.~\ref{SSggcoup}.

However, this is problematic: calculations of these diagrams can only
be done in unitary gauge, and here there are serious divergences.
Fortunately, in this case there is a solution.

First, consider the coupling of two $U_1$ LQs to the hypercharge boson
$B_\mu$. From Eq.~(\ref{U1U1terms}), this is
\bea
{\cal L}_{U_1U_1B} &=& i g' \frac23 \left[ (\partial_\mu U^\dagger_{1\nu} - \partial_\nu U^\dagger_{1\mu} ) B^\mu U_1^\nu
  - (\partial_\mu U_{1\nu} - \partial_\nu U_{1\mu} ) B^\mu U_1^{^\dagger\nu} \right. \nn\\
  && \hskip3truecm \left.  +~\frac{g_{U_1U_1B}}{g'}
  ( U_{1\mu} U^\dagger_{1\nu} - U_{1\nu} U^\dagger_{1\mu} ) \partial^\mu B^\nu \right] ~.
\label{gaugecouplings}
\eea
From the point of view of the effective Lagrangian below
$\Lambda_{NP}$, $g_{U_1U_1B}$ is an arbitrary constant. However, one
must remember that both $U_1$ and $B_\mu$ are gauge bosons of the
(unspecified) larger gauge group. As such, their interactions arise
only from the gauge-boson kinetic term $F^i_{\mu\nu} F^{i\mu\nu}$ of
this gauge group. This will necessarily generate the three terms above
with equal weighting, i.e., $g_{U_1U_1B} = g'$, as occurs in the SM
with the $W^+W^-\gamma$ coupling terms. (See
Ref.~\cite{Biggio:2016wyy} for similar arguments in the context of
$(g-2)_\mu$.)  The coupling of two $U_1$ LQs to gluons can be treated
similarly, leading to $g_{U_1U_1g} = g_s$.

Second, note that the diagrams in Fig.~\ref{SSggcoup} are essentially
the same as those of the SM process $h \to \gamma\gamma$ with internal
$W$ bosons, with the obvious changes of $h$ replaced by $SS$, $W$ by
$U_1$, and $\gamma$ by $g$. As discussed above, the $U_1 {\bar U}_1 g$
and $U_1 {\bar U}_1 g g$ couplings have the same structure as the SM
$W^+ W^- \gamma$ and $W^+ W^- \gamma \gamma$ couplings. In
Ref.~\cite{Marciano:2011gm}, the SM $h \to \gamma\gamma$ diagrams with
internal $W$s are calculated in unitary gauge, and it is explicitly
shown that the result is finite and gauge-independent.

A more general calculation can be found in
Ref.~\cite{Belanger:2013oya}. There it is found that, for the
effective coupling term $\lambda h F_{\mu \nu} F^{\mu \nu}$, the
contribution to $\lambda$ from $g_{hVV} M_V h V_\mu V^\mu $ is given
by
\beq
\lambda = -\frac{\alpha}{8\pi} \frac{g_{hVV} f_V^c q_V^2}{2 M_V} A({\tau}) ~~,~~~~
\tau=\frac{M_h^2}{4M_V^2} ~.
\label{Mhfactor}
\eeq
Here $q_V$ and $f_V^c$ are the electric charge and the colour factor of
the vector particle running in the loop. (For the $W$, $q_V = f_V^c =
1$.) The loop function is given by
\bea
A(x) &=& - \frac{1}{x^2} [ 2x^2 + 3x + 3(2x-1) f(x) ] ~, \nn\\
{\rm with} ~~~~~ f(x) &=&
\begin{cases}
  \arcsin^2(\sqrt{x}) & {\rm for}~x\le 1 ~, \\
  -\frac14 \left[ \ln \frac{1 + \sqrt{1-x^{-1}}}{1 - \sqrt{1-x^{-1}}} - i\pi \right]^2 & {\rm for}~x > 1 ~. \\ 
\end{cases}
\eea
In the limit of $x \to 0$, $A(x) \to -7$ \cite{Ellis:1975ap}.

From this we can deduce the coefficient of the effective coupling term
$\alpha_s C_{5}^g S^2 G_{\mu \nu}^a G^{a \mu \nu}$ generated by the
coupling $- \frac12 \lambda_{U_1 S} \, S^2 \, U_{1 \mu}^{\dagger}
U^{\mu}_1$. We make the substitutions $\alpha \to \alpha_s$, $g_{hVV}
\to \frac12 \lambda_{U_1 S}$ and $f_V^c = \rm{Tr}(T^a T^a)=\frac12$
($q_V$ is not relevant). Also, in Eq.~(\ref{Mhfactor}), $M_h^2 \to 2
k_1 \cdot k_2$, where the $k_i$ are the four-momenta of the
gluons. Thus, this factor is process-dependent. We then have
\beq
C_{5}^g = -\frac{\lambda_{U_1 S}}{64 \pi M_{U_1}^2} A({\tau}) ~~,~~~~
\tau=\frac{k_1 \cdot k_2}{2 M_{U_1}^2} ~.
\label{C5g}
\eeq
The scattering process $S N \to S N$ takes place at energies of
$O({\rm keV})$, so it is reasonable to consider the limit $\tau \to
0$. For DM annihilation in the early universe via $S S \to g g$, we
have $k_1 \cdot k_2 \simeq 2 M_S^2$.

The gluons in Fig.~\ref{SSggcoup} can be replaced by photons, leading
to an $SS\gamma\gamma$ coupling.  Writing the effective coupling term
as $\alpha C_{5}^\gamma S^2 F_{\mu \nu} F^{\mu \nu}$, and using $q_{U_1} =
\frac23$ and $f^c_{U_1} = 3$, we have from Eq.~(\ref{Mhfactor}) that
\beq
C_{5}^\gamma = -\frac{\lambda_{U_1 S}}{48 \pi M_{U_1}^2} A({\tau}) ~~,~~~~
\tau=\frac{k_1 \cdot k_2}{M_{U_1}^2} ~.
\eeq
The $SS \to \gamma\gamma$ annihilation channel can be important for
indirect detection, as we will discuss in Sec.~\ref{DMObservables}.

\begin{figure}[!htbp]
\begin{center}
\includegraphics[width=0.40\textwidth]{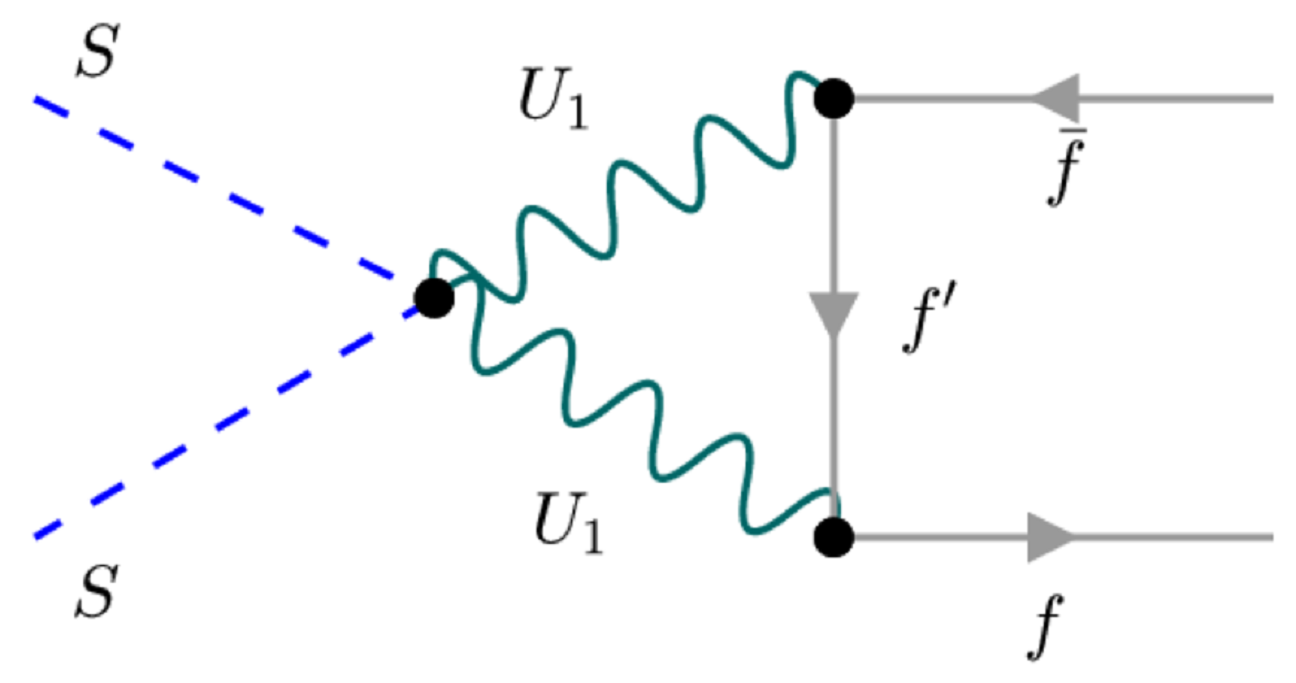}
\end{center}
\vskip-5truemm
\caption{\small Diagram generating the $SS{\bar f}f$ coupling. If $f$
  is a quark, $f'$ is a lepton and vice-versa.}
\label{SSffcoup}
\end{figure}

Finally, another coupling that can be generated at one loop is
$SS{\bar f}f$, where $f$ is any SM fermion, see Fig.~\ref{SSffcoup}.
However, in unitary gauge, this diagram diverges, so this coupling
cannot be computed in the absence of a complete model. Even so, as we
see below, a simple dimensional analysis is sufficient.

First, we note that it is usually assumed that the $U_1$ couples only
to third-generation fermions in the gauge basis. The (smaller)
couplings involving the second generation are generated when one
transforms to the mass basis. In Ref.~\cite{Kumar:2018kmr}, it was
shown that the $B$ anomalies can be explained if the couplings are
purely left-handed with
\beq
g_{L}^{33} = 1 ~~,~~~~ g_{L}^{23} = g_{L}^{32} = 0.28 ~~,~~~~ g_{L}^{22} = -0.008 ~,
\label{U1couplings}
\eeq
where the $g_L^{ij}$ are the Yukawa couplings of
Eq.~(\ref{SSU1U1coup}), with $g_R^{ij} = 0$. The $SS{\bar f}f$
coupling then takes the form
\beq
SS{\bar f}f \sim \frac{\lambda_{U_1 S} |g_L^{ij}|^2 }{16 \pi^2 M_{U_1}^2} 
\left[ m_f \, SS{\bar f}P_{L}f
  + S {\lrvec{\partial}}_\mu S {\bar f}\gamma^\mu P_{L}f \right] ~,
\label{SSffcoupling}
\eeq
where the fermion $f$ can be a quark $q_i$ or a lepton $l_j$.

We can use this to estimate the contribution to the relic density from
the $SS{\bar f}f$ operator. The coefficients of the first and second
terms are $O(10^{-7})~{\rm GeV}^{-1}$ (for $f = t$) and
$O(10^{-9})~{\rm GeV}^{-2}$, respectively. Both of these $SS{\bar f}f$
couplings are examined in Ref.~\cite{Beltran:2008xg}, and as can be
seen from Fig.~5 of this paper, these coefficients are far too small
to generate an important contribution to the relic density.

If $f$ is a quark, the $SS{\bar f}f$ coupling could in principle
contribute to $S N \to S N$ (direct detection). However, we see that
this contribution is small: the Yukawa couplings are large only for
third-generation quarks, whose presence in nucleons is suppressed.
Conversely, first-generation quarks are plentiful in nucleons, but the
Yukawa couplings are tiny. The bottom line is that this contribution
to $S N \to S N$ is negligible compared to that of $SSgg$.

\subsection{Scalar Singlet Dark Matter Model}

The scalar DM couples to the Higgs via the Higgs portal term in
Eq.~(\ref{portals}), $S^2 |H|^2$. This coupling is the basis for the
SSDMM \cite{Silveira:1985rk, McDonald:1993ex, Burgess:2000yq}.

After electroweak symmetry breaking, $H \to (h + v_0)/\sqrt{2}$, where
$h$ is the physical Higgs boson and $v_0 = 246$ GeV is the VEV of the
Higgs field. Three terms are generated from the Higgs portal term:
\beq
- \frac14 \lambda_{hS} v_0^2 S^2 - \frac12 \lambda_{hS} v_0 S^2 h - \frac14 \lambda_{hS} S^2 h^2 ~.
\label{lambdahSterms}
\eeq
The terms $v_0 S^2 h$ and $S^2 h^2$ lead to DM annihilation via the
processes $S S \to h \to W^+ W^-$, $Z^0 Z^0$, $h h$, $t {\bar t}$, $b
{\bar b}$, $g g$, etc., and $SS \to hh$, respectively.  Given the relic
density of DM, and assuming that the $S$ particles were in thermal
equilibrium with ordinary matter throughout the evolution of the
universe, the present-day value of the density of DM
\cite{Planck:2018vyg} fixes the value of $\lambda_{hS}$.

The $v_0 h S^2$ term has other effects. It leads to elastic scattering
of $S$ with nuclei via Higgs exchange: $S N \to S N$. Direct detection
experiments look for this process. In addition, if the $S$ is
sufficiently light -- its mass is $M_S = \sqrt{\mu_S^2 + \frac12
  \lambda_{hS} v_0^2}$ -- the invisible decay of the Higgs, $h \to S
S$, is possible. Finally, this term can also generate $SS \to
\gamma\gamma$, a process that is relevant for indirect detection. It
is therefore possible in principle that the SSDMM can account for all
DM observations without the addition of any additional NP.

In 2017, a global fit of the SSDMM was performed, taking into account
the measurements of the relic density, direct and indirect detection,
and the invisible Higgs width \cite{GAMBIT:2017gge}. (Other analyses
of the SSDMM can be found in Refs.~\cite{Cheung:2012xb, Cline:2013gha,
  Beniwal:2015sdl, Athron:2018ipf}.) It was found that the value of
$\lambda_{hS}$ required to satisfy the relic density constraint
implies a DM scattering rate on nuclei that exceeds the Xenon1T limits
when $M_S \lsim 500$-700 GeV, except in some small (fine-tuned)
regions of parameter space where $M_S\simeq 60$ GeV. And while the
model is still viable for large values of $M_S$, for $M_S \gsim 5$
TeV, the required value of $\lambda_{hS}$ becomes large, approaching
the nonperturbative regime.

Note that Ref.~\cite{GAMBIT:2017gge} uses the 2017 direct detection
constraints of the Xenon1T Collaboration \cite{Aprile:2017iyp} in its
analysis of the SSDMM. These constraints have been improved recently
(2021) by the PandaX-4T Collaboration \cite{PandaX-4T:2021bab}. While
we do not perform a complete analysis of the SSDMM in this paper, in
the next section we do show how these more-stringent direct detection
constraints affect this model.

\section{Effect on DM Observables}
\label{DMObservables}
  
The most direct consequence of an $S^2 \, U_{1 \mu}^{\dagger}
U^{\mu}_1$ coupling is the generation of the tree-level annihilation
process $S S \to U_1 {\bar U}_1$. It is well known that the process of
DM annihilation takes place mostly at very small relative momentum, so
that this process essentially does not occur for $M_S < M_{U_1} = 1.8$
TeV.  In examining the effects of the $S^2 \, U_{1 \mu}^{\dagger}
U^{\mu}_1$ coupling on the various DM observables, it is therefore
useful to consider separately heavy DM ($M_S \gsim M_{U_1} = 1.8$ TeV)
and light DM ($M_S \lsim M_{U_1} = 1.8$ TeV). In our analysis of the
$U_1$ contributions to DM processes, we also include the Higgs-portal
contributions.

Our DM analysis is performed using two different programs (to provide
crosschecks): MadDM \cite{Ambrogi:2018jqj} and micrOMEGAs
\cite{Belanger:2008sj, Belanger:2018ccd}. The results for heavy and
light DM presented below use MadDM and micrOMEGAs, respectively
(unless otherwise indicated).

\subsection{\boldmath Heavy DM ($M_S \gsim M_{U_1} = 1.8$ TeV)}
\label{heavyDM}

\subsubsection{Relic Density}
\label{heavyMS_RD}

We begin by considering $M_S \gsim M_{U_1} = 1.8$ TeV. For this heavy
DM, the lowest-order contribution of the $U_1$ portal to the relic
density, namely $S S \to U_1 {\bar U}_1$, is possible.

As pointed out above, the $S^2 \, U_{1 \mu}^{\dagger} U^{\mu}_1$
coupling will generate at one loop a $C_{5}^g S^2 G_{\mu \nu}^a G^{a
  \mu \nu}$ term [Eq.~(\ref{C5g})] and an $SS{\bar f}f$ coupling
[Eq.~(\ref{SSffcoupling})].  These will lead to contribution to the
relic density from the annihilations $SS \to gg$ and $SS \to f{\bar
  f}$. However, because they are essentially loop-level processes,
they are negligible compared to $S S \to U_1 {\bar U}_1$.

DM annihilation takes place via both the Higgs and $U_1$ portals,
whose couplings are respectively $\lambda_{hS}$ and $\lambda_{U_1 S}$.
To calculate the constraints from the relic density, we use MadDM
\cite{Ambrogi:2018jqj}.  We fix $M_{U_1} = 1.8$ TeV and take
$\lambda_{U_1 S} = 0$ (pure Higgs portal), 0.04, 0.07 and 0.10.  For
each value of $\lambda_{U_1 S}$, we compute the value of
$\lambda_{hS}$ required to reproduce the relic density $\Omega h^2 =
0.1198 \pm 0.0012$ \cite{Planck:2018vyg} to within $\pm 3\sigma$.

\begin{figure}[!htbp]
\begin{center}
\includegraphics[width=0.70\textwidth]{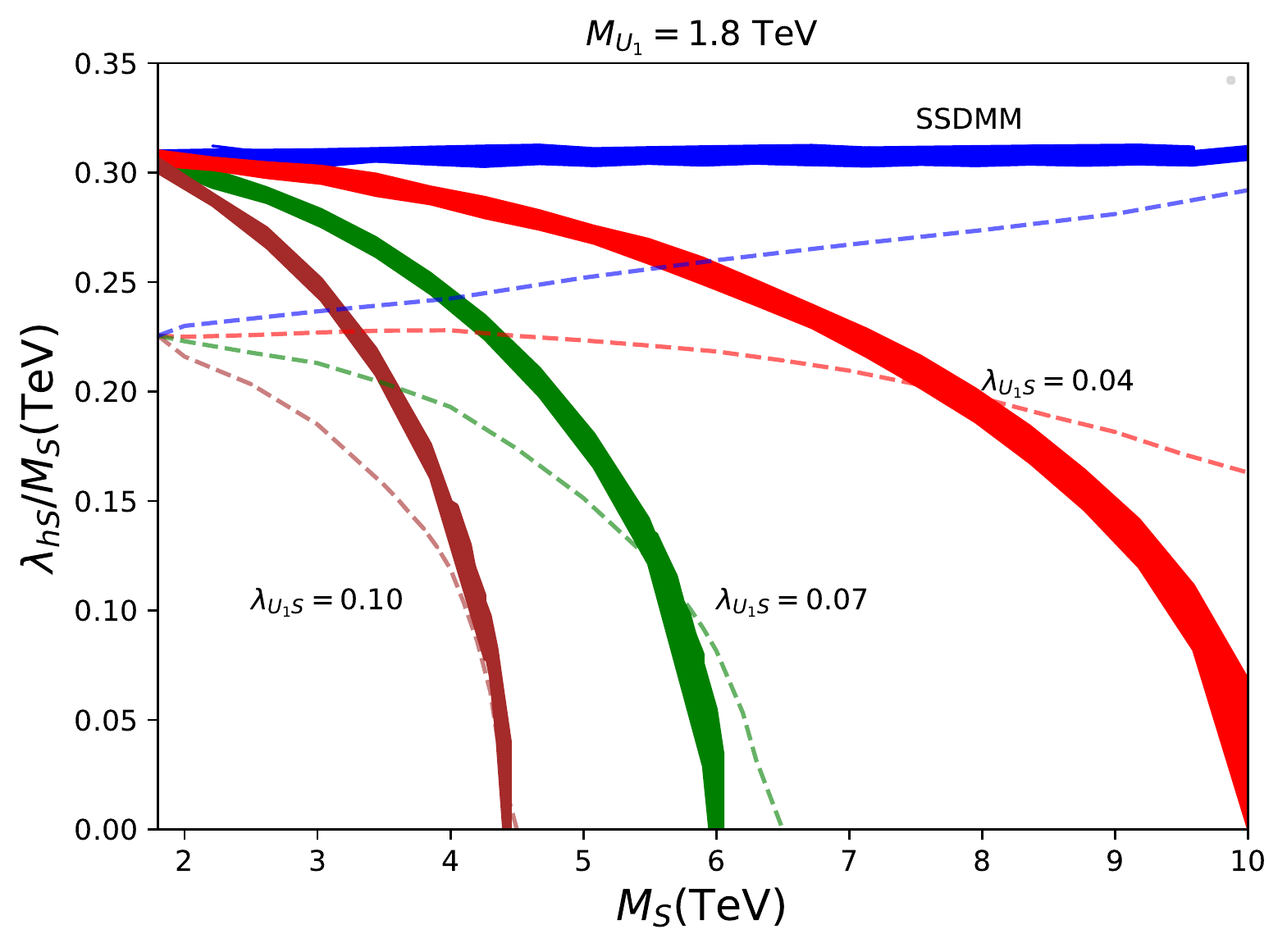}
\end{center}
\caption{\small Value of $\lambda_{h S}/M_S({\rm TeV})$ required to
  reproduce the relic density in the SSDMM (blue) and in the modified
  SSDMM with $\lambda_{U_1 S} = 0.10$ (brown), 0.07 (green) and 0.04
  (red), for $M_{U_1} = 1.8$ TeV. all as a function of $M_S$.  Dashed
  lines indicate the future limits from indirect detection searches by
  CTA \cite{CTA:2020qlo}.}
\label{relic}
\end{figure}

The results are shown in Fig.~\ref{relic}. We note the following:
\begin{itemize}

\item For $\lambda_{U_1 S} = 0$ (pure Higgs portal), the relic density
  can be reproduced for all values of $M_S$ in the range 1 TeV $\le
  M_S \le$ 10 TeV. However, for $M_S \gsim 5$ TeV, the required value
  of the coupling $\lambda_{hS}$ becomes large, approaching
  nonperturbative values.

\item The addition of a nonzero $\lambda_{U_1 S}$ opens up the $U_1$
  portal, so that the Higgs portal does not have to reproduce the
  relic density by itself. As a result, the required value of
  $\lambda_{hS}$ is decreased. The size of the effect of the $U_1$
  portal depends on the value of $\lambda_{U_1 S}$.

\item Consider $\lambda_{U_1 S} = 0.10$. For $M_S \simeq 2$ TeV, the
  required value of $\lambda_{hS}$ is smaller than for the case of a
  pure Higgs portal. As $M_S$ increases, the rate for DM annihilation
  into $U_1 {\bar U}_1$ also increases, until at $M_S \simeq 4.5$ TeV,
  the Higgs portal is not even required ($\lambda_{hS} = 0$). But in
  this case, for even larger values of $M_S$, the annihilation rate
  into $U_1 {\bar U}_1$ is too large, so that the relic density is
  always below the measured value.

\item One sees this same type of behaviour for smaller values of
  $\lambda_{U_1 S}$.  For $\lambda_{U_1 S} = 0.07$, the maximal value
  of $M_S$ is $\simeq 6$ TeV, while for $\lambda_{U_1 S} = 0.04$ it is
  between 10 and 11 TeV.

\item Similarly, for each value of $M_S$, there is a value of
  $\lambda_{U_1 S}$ above which the relic density is too low. For
  example, for $M_S = 3$ TeV, $\lambda_{U_1 S} < 0.17$ is required.
  
\item In the presence of the $U_1$ portal, the required value of
  $\lambda_{hS}$ for $M_S \gsim 5$ TeV remains small and clearly
  perturbative, provided $\lambda_{U_1S}$ is large enough, in contrast
  to the case of the pure Higgs portal.

\end{itemize}

\subsubsection{Indirect Detection}
\label{HeavyDM_ID}

DM annihilation in galaxies also provides a method for indirect
detection using various final states. Here we consider final states
that contain photons. There are two different categories of such final
states. First, there is the loop-induced process $SS\to \gamma\gamma$,
which gives rise to a characteristic monoenergetic photon line at an
energy of $M_S$. Second, the photons can be emitted from other
final-state particles. Taking into account DM annihilation into all
final states, this results in a continuous photon spectrum.

Photons from $SS\to \gamma\gamma$ were searched for by the
H.E.S.S.\ Collaboration in observations from the inner Galactic halo
\cite{HESS:2018cbt}.  However, the limits placed on the cross section
are several orders of magnitude larger than what is predicted in our
model. This is because, for heavy DM, $\lambda_{U_1 S}$ is much
smaller than unity, leading to a significant suppression of this
loop-induced process. H.E.S.S.\ also measured the continuous photon
spectrum, placing constraints on the DM annihilation cross-section
$\langle \sigma v \rangle \approx 10^{-24} ~\rm{cm}^3/s$
\cite{Fermi-LAT:2019lyf}. This is two orders of magnitude above the
typical thermal cross-section of ${\cal O}(10^{-26} \rm{cm}^3/s)$
found in our model.

More useful limits on DM annihilation will be obtained by the
Cherenkov Telescope Array (CTA), which measures the continuous
photon spectrum at higher energies \cite{CTA:2020qlo}, and is
therefore relevant for TeV-scale DM.

Using micrOMEGAs\_5.2, we compute the continuous photon spectrum for
heavy DM annihilation into all final states.  This value is then
compared with the one that can be probed by CTA in the Galactic Center
using the projections given in Ref.~\cite{CTA:2020qlo} and assuming an
Einasto DM profile.  To do this comparison, we use the likelihood
tables provided in Ref.~\cite{zenodo}. When $\lambda_{U_1S}=0$, the
dominant SM final states are $WW$, $ZZ$ and $hh$. In this case, we
find that CTA will be able to probe the full region of the parameter
space that leads to the correct relic density. This corresponds
roughly to $\lambda_{hS}\simeq 0.25 M_S$, and will rule out the SSDMM
for heavy DM if no signal is observed.

As $\lambda_{U_1S}$ increases such that the $SS \to U_1 \bar{U}_1$
contribution becomes more important, smaller values of $\lambda_{hS}$
can be probed. Nevertheless, for each value of $\lambda_{U_1S}$ shown
in Fig.~\ref{relic}, when the DM mass exceeds a certain value, the
relic-density-favoured region cannot be probed (e.g., for
$\lambda_{U_1S} = 0.10$, this value is $\simeq 4.5$ TeV). In
Sec.~\ref{heavyMS_RD}, we pointed out that, for each value of
$\lambda_{U_1S} > 0$, there is a value of $M_S$ beyond which the DM
annihilation channel into $U_1 \bar{U_1}$ becomes too efficient, and
the relic density condition can never be satisfied. Similarly, there
is a value of $M_S$ for which DM annihilation in the galaxy is totally
dominated by the $U_1$ final state and the CTA reach is independent of
$\lambda_{hS}$. For example, for $\lambda_{U_1S} = 0.10$ (0.07), this
value is $\simeq 4.5$ TeV ($\simeq 6.5$ TeV). Hence we conclude that,
for $\lambda_{U_1S} = 0.10$, the model can be completely probed by CTA,
while for smaller values of $\lambda_{U_1S}$, larger values of $M_S$
are beyond the reach of CTA.

Note that when DM annihilates primarily into $U_1 {\bar U}_1$, the
limit extracted from indirect detection shows little dependence on the
$U_1$ couplings to fermions. For example, variations of $\pm 20\%$ in
the couplings of Eq.~(\ref{U1couplings}) change the indirect detection
bounds by less than 1\%.

\subsubsection{Direct Detection}

In Sec.~\ref{heavyMS_RD}, we saw that the addition of the $U_1$ portal
has the effect of reducing the required value of $\lambda_{hS}$.  But
this has a downside as well. $\lambda_{hS}$ is also the coefficient of
the $v_0 h S^2$ term in Eq.~(\ref{lambdahSterms}), which leads to
processes observable in direct detection experiments. Thus, a smaller
value of $\lambda_{hS}$ implies a decreased direct detection
signal. In principle, this could be compensated for by direct
detection signals using the $U_1$ LQ.  Unfortunately, at tree level,
there are no such signals.  As a result, the addition of the $U_1$
portal has the effect of reducing the direct detection signal.

\begin{figure}[!htbp]
\begin{center}
\includegraphics[width=0.70\textwidth]{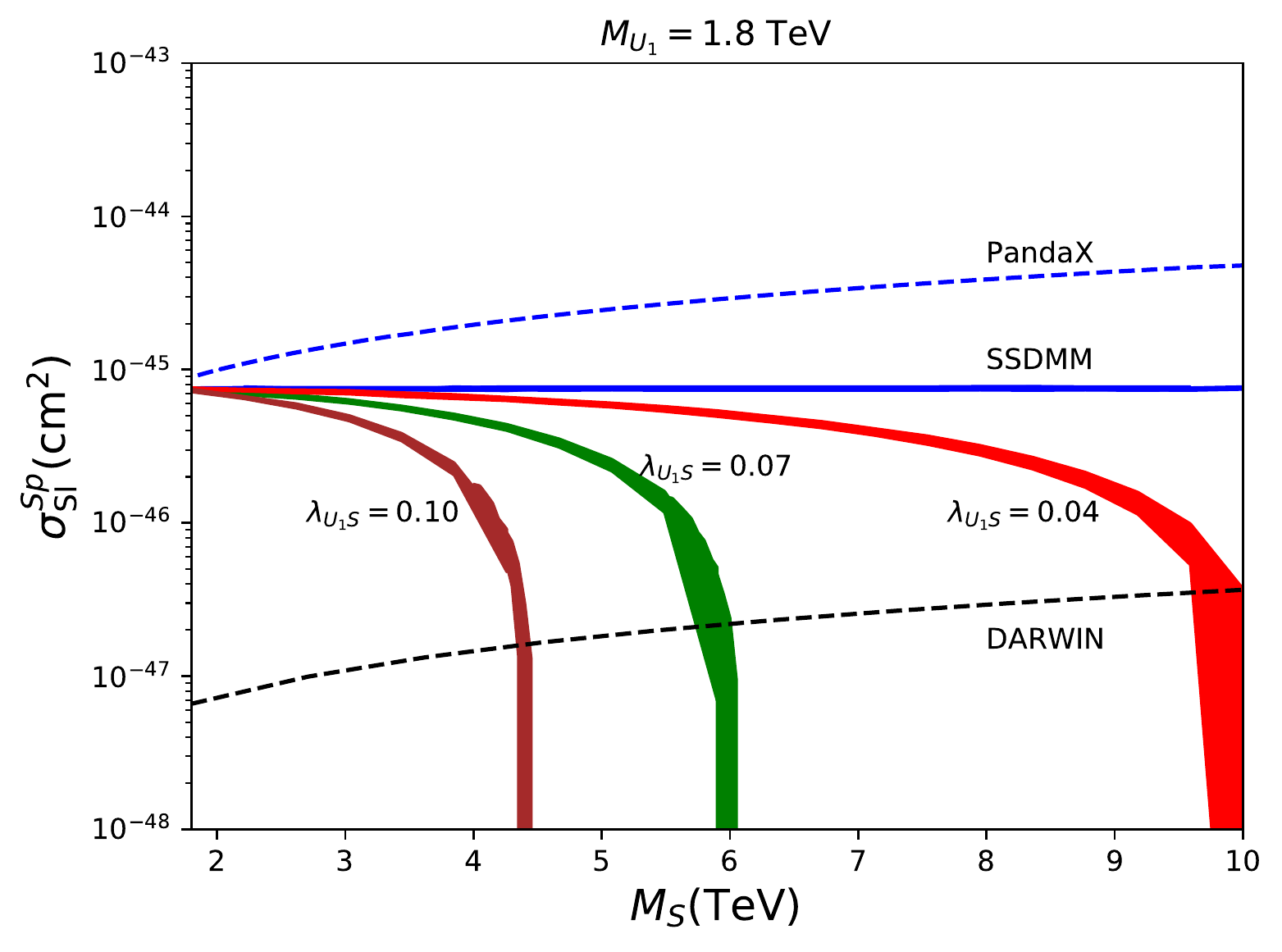}
\end{center}
\caption{\small Predicted spin-independent cross section $\sigma_{{\rm
      SI}}^{Sp}$ in the SSDMM (blue) and in the modified SSDMM with
  $\lambda_{U_1 S} = 0.10$ (brown), 0.07 (green) and 0.04 (red), for
  $M_{U_1} = 1.8$ TeV. all as a function of $M_S$. The dashed lines
  are the present upper limit on the cross section from the PandaX-4T
  Collaboration \cite{PandaX-4T:2021bab} (blue) and the future reach
  of the DARWIN Collaboration \cite{DARWIN:2016hyl} (black).}
\label{dd}
\end{figure}

This is illustrated in Fig.~\ref{dd}, where we present the results for
the spin-independent cross section for DM scattering off protons,
$\sigma_{{\rm SI}}^{Sp}$.  If only the Higgs portal is available
($\lambda_{U_1 S} = 0$), the SSDMM yields a cross section
$\sigma_{{\rm SI}}^{Sp} \simeq 10^{-45}~{\rm cm}^2$.  If the $U_1$
portal is also open ($\lambda_{U_1 S} \ne 0$), this cross section
decreases. The size of the decrease depends on how much $\lambda_{h
  S}$ is reduced, which itself depends on the values of $\lambda_{U_1
  S}$ and $M_S$, following the pattern of Fig.~\ref{relic}.

Note that there is also a contribution to $S p \to S p$ scattering due
to the one-loop $SSgg$ coupling [Eq.~(\ref{C5g})]. However, for heavy
DM, this contribution is negligible compared to that of the Higgs. (In
the next subsection, we will see that this is not the case for light
DM.)

At the present time, the upper limit on $\sigma_{{\rm SI}}^{Sp}$ from
the PandaX-4T Collaboration is between $10^{-45}~{\rm cm}^2$ and
$10^{-44}~{\rm cm}^2$, for 1 TeV $\le M_S \le$ 10 TeV
\cite{PandaX-4T:2021bab}. As can be seen from Fig.~\ref{dd}, this is
still larger than the values predicted for the cross section in the
SSDMM and the modified SSDMM, which includes the $U_1$ LQ.  In the
future, XENONnT \cite{XENON:2015gkh} and DARWIN \cite{DARWIN:2016hyl}
will improve these constraints by almost two orders of magnitude.  If
no signal is observed, this will rule out the SSDMM. However, the
modified SSDMM will still be viable for certain values of
$\lambda_{U_1 S}$ and $M_S$.

\subsection{\boldmath Light DM ($M_S < M_{U_1} = 1.8$ TeV)}
\label{lightDM}

For DM of mass $M_S < 1.8$ TeV, annihilation via $S S \to U_1
{\bar U}_1$ is not possible. Instead, one must rely on the SSDMM.
However, as shown in Ref.~\cite{GAMBIT:2017gge}, this is
problematic. For some values of $M_S$ in this mass range, the value of
the $SSh$ coupling $\lambda_{hS}$ required to reproduce the relic
density via $S S \to h \to {\rm SM~particles}$ is in tension with the
constraints from direct detection.

The one-loop $SSgg$ coupling generated via a virtual $U_1$ has the
potential to help. First, it provides another annihilation channel,
$SS \to gg$, so that the required value of $\lambda_{hS}$ can be
reduced. Second, it also contributes to the process $S p \to S p$ used
for direct detection. It is possible that this contribution interferes
destructively with that of the SSDMM, leading to weaker constraints on
$\lambda_{hS}$ from direct detection.

In addition, there is the possibility that one of the final-state LQs
in $S S \to U_1 {\bar U}_1$ is virtual, leading to the three-body
annihilation $S S \to U_1 {\bar q} \ell$. This could be important for
$M_S > M_{U_1}/2$. To perform this calculation, we use the preferred
couplings of the $U_1$ to ${\bar q} \ell$ given in
Eq.~(\ref{U1couplings}). We include the contribution of the three-body
annihilation channels in micrOMEGAs\_5.2 as follows.  We use the
feature of micrOMEGAs that allows one to substitute a new annihilation
cross section in the relic density calculation. We compute the
cross-sections for all relevant three-body processes (namely $U_1 \tau
b$, $U_1 \nu_\tau t$, $U_1 \tau s$, $U_1 \nu_\tau c$). From this we
subtract $SS\rightarrow U_1 {\bar U}_1$ to avoid double counting when
$M_S$ is near $M_{U_1}$ and the $U_1 {\bar U}_1$ can be produced on
shell. This is then added to the two-body processes.

\begin{figure}[!htbp]
\begin{center}
\includegraphics[width=0.70\textwidth]{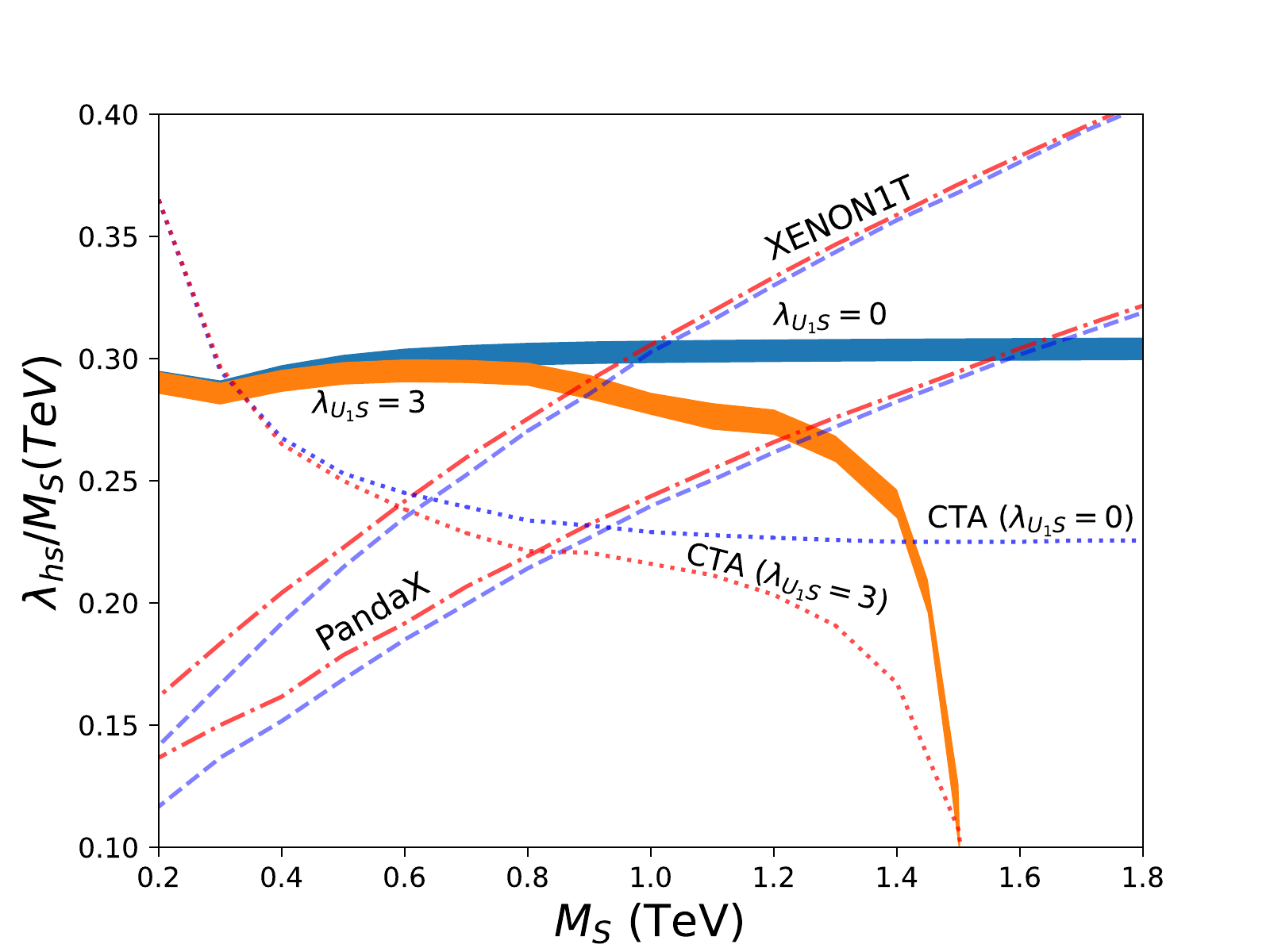}
\end{center}
\caption{\small Values of $\lambda_{hS}/M_S({\rm TeV})$ required to
  reproduce the relic density (within $\pm 3\sigma$) as a function of
  $M_S$, for $\lambda_{U_1 S} = 0$ (blue band) and $\lambda_{U_1 S} =
  3.0$ (orange band). Dashed lines show present constraints from the
  upper limit on the spin-independent cross section $\sigma_{{\rm
      SI}}^{Sp}$ \cite{Aprile:2017iyp} for the SSDMM
  (blue) and the SSDMM + $U_1$ (orange). Dotted lines indicate the
  future limits from indirect detection searches by CTA
  \cite{CTA:2020qlo}.}
\label{DD_lightDM}
\end{figure}

The consequence of all of these effects for the SSDMM is illustrated
in Fig.~\ref{DD_lightDM}. Here we compute the value of
$\lambda_{hS}/M_S({\rm TeV})$ required to reproduce the relic density
(within $\pm 3\sigma$) as a function of $M_S$. We overlay the
constraints\footnote{In order to compute $\sigma_{{\rm SI}}^{Sp}$, we
  use the default values of micrOMEGAs for the coefficients of the
  quark content in the nucleon \cite{Belanger:2013oya}. However, the
  $SSgg$ operator is ignored when computing the direct interaction
  rate. To simulate the effect of this operator, we introduce a new
  heavy quark and a new heavy scalar mediator ($H_S$) in the model,
  see Ref.~\cite{Belanger:2008sj}. The mediator couples only to the
  heavy quark and to DM as ${\cal L}= -m_Q \bar{Q} {Q} H_S + x SS
  H_S$, where $x=21/16 \lambda_{U_1S} M_{H_S}^2/M_{U1}^2$.} from the
Xenon1T Collaboration \cite{Aprile:2017iyp} (2017) and the PandaX-4T
Collaboration \cite{PandaX-4T:2021bab} (2021) due to their upper limit
on the spin-independent cross section $\sigma_{{\rm SI}}^{Sp}$.

For the case of $\lambda_{U_1 S} = 0$ (pure Higgs portal), we see
that, using the 2017 direct detection constraints from the Xenon1T
Collaboration \cite{Aprile:2017iyp}, there is a solution only for $M_S
\gsim 950$ GeV. This explicitly demonstrates the problems for the
SSDMM that were found in Ref.~\cite{GAMBIT:2017gge}\footnote{Note that
  this restriction on $M_S$ is not exactly the same as that found in
  Ref.~\cite{GAMBIT:2017gge}. This is because the analyses are not the
  same -- we are simply taking into account constraints from the relic
  density and direct detection, while Ref.~\cite{GAMBIT:2017gge}
  performs a more complete global analysis. Still, the point is that
  the SSDMM had difficulties reproducing the data even in 2017.}. But
when the 2021 direct detection constraints from the PandaX-4T
Collaboration \cite{PandaX-4T:2021bab} are used, one sees that these
problems are worse: now, the SSDMM is viable only for $M_S \gsim 1.6$
TeV.

As expected, the situation improves when one adds the $U_1$ portal
term $S^2 \, U_{1 \mu}^{\dagger} U^{\mu}_1$. To be specific, with a
(large) value of $\lambda_{hS} = 3.0$, the minimum value of $M_S$ for
which an explanation of the relic density can be found is reduced from
$\simeq 1.6$ TeV to $\simeq 1.3$ TeV. There are several reasons for
this. First, there are additional annihilation channels. $SS \to gg$
is possible for all values of $M_S$. And for $M_S > M_{U_1}/2$, the
three-body annihilation $S S \to U_1 {\bar q} \ell$ opens up: the
required value of $\lambda_{hS}/M_S({\rm TeV})$ begins to diminish at
$M_S \simeq 1$ TeV, and falls preciptously at $M_S \simeq 1.65$ TeV.
Second, the destructive interference of the $SSgg$ and Higgs
contributions to $S p \to S p$ plays a (smaller) role.

We therefore conclude that, for light DM, the addition of the $U_1$ LQ
does somewhat improve the outlook for the SSDMM (though not to the
extent of allowing an explanation of the DM observables for $M_S =
O(100)$ GeV).

As for indirect detection, the present bounds on $\sigma(SS\to
\gamma\gamma)$ \cite{HESS:2018cbt, Fermi-LAT:2015kyq} are several
orders of magnitude larger than the prediction of our model.  Even for
$\lambda_{U_1S}=3$, which leads to the maximal effect, we obtain $v
\sigma_{\gamma\gamma}= 5.5\times 10^{-31}~{\rm cm}^3/{\rm s}$ for
$M_S=400$ GeV, while the bounds from FermiLAT (H.E.S.S.) are $4~(2.8)
\times 10^{-28}~{\rm cm}^3/{\rm s}$, assuming the Einasto
profile. Similarly, for $M_S=1000$ GeV, we predict $v
\sigma_{\gamma\gamma}= 9.7 \times 10^{-30}~{\rm cm}^3/{\rm s}$, while
the bound from H.E.S.S.\ is $3.5 \times 10^{-28}~{\rm cm}^3/{\rm s}$.
From measurements of the continuous photon spectrum, FermiLAT can
probe the thermal cross section only for DM masses around 100 GeV
\cite{Fermi-LAT:2015att}. However, future measurements of the
continuous photon spectrum from DM annihilation by CTA in the Galactic
Center will be able to rule out the SSDMM and place stringent
constraints on the modified SSDMM with $\lambda_{U_1 S} \ne 0$, as can
be seen in Fig.~\ref{DD_lightDM}.

Finally, an $S^2 \, U_{1 \mu}^{\dagger} U^{\mu}_1$ coupling will also
lead to the LHC scattering process $p p \to \ell^+ \ell^- S S$ via the
$t$-channel exchange of a $U_1$ LQ. Here the $SS$ would of course be
``observed'' as missing energy. We have computed the cross section for
this process for $M_S = 100$ GeV. We find that, because the $U_1$ LQ
is so heavy, and because it couples mainly to third-generation quarks
[see Eq.~(\ref{U1couplings})], the cross section is tiny: for
$\lambda_{U_1 S} = O(1)$, $\sigma(p p \to \ell^+ \ell^- S S) =
O(10^{-6}$ pb, which is unobservable.

\section{Conclusions}

At present, there are several $B$-decay observables whose measured
values exhibit discrepancies with the predictions of the SM. These
decays are mediated by $\bsmumu$ or $\bctaunu$ transitions. It is
possible to find NP models that explain both types of $B$ anomalies.
When all constraints are taken into account, only one model survives.
It involves the addition of the $U_1$ LQ, a vector particle of charge
$2/3$ that is an $SU(2)_L$ singlet and has a mass $M_{U_1} \gsim 1.8$
TeV.

Another observation that is unexplained by the SM is DM. A
particularly simple scenario to explain the observed relic density is
the scalar singlet dark matter model (SSDMM). We update the
constraints on the SSDMM, taking into account the 2021
direct detection constraints from the PandaX-4T Collaboration. We find
that this model is now viable only for $M_S \gsim 1.6$ TeV.
Furthermore, for $M_S = O({\rm TeV})$, the required value of
$\lambda_{hS}$ enters the nonperturbative regime.

In an attempt to improve the prospects for the SSDMM, we add the $U_1$
LQ and assume a $\lambda_{U_1 S} \, S^2 \, U_{1 \mu}^{\dagger}
U^{\mu}_1$ coupling term. For heavy DM ($M_S \gsim M_{U_1}$), such a
coupling leads to DM annihilation via $S S \to U_1 {\bar U}_1$. This
coupling will also lead to $SSgg$ and $SS\gamma\gamma$ couplings at
one loop. Although naively divergent, these loop diagrams can be
computed by analogy to the SM $hgg$ or $h\gamma\gamma$ couplings,
under the assumption that the $U_1$ is a gauge boson of a group broken
at the TeV scale. The $SSgg$ and $SS\gamma\gamma$ couplings provide
additional annihilation channels for light DM ($M_S < M_{U_1}$), and
lead to signals of direct and indirect detection.

For heavy DM, the addition of the $S S \to U_1 {\bar U}_1$
annihilation channel does indeed reduce the required value of
$\lambda_{hS}$ to below the nonperturbative level. The downside of
this is that the direct detection signal is also reduced. Still,
future direct detection experiments will be able to place constraints
that will rule out the SSDMM, but the modified SSDMM will still be
viable for certain values of $\lambda_{U_1 S}$ and $M_S$. And future
indirect detection measurements will be able to place stringent
constraints on the modified SSDMM.

For light DM, the $SSgg$ coupling provides another annihilation
channel, and yields a contribution to direct detection that can
interferes destructively with that of the SSDMM.  And for $M_S >
M_{U_1}/2$, one can have $S S \to U_1 {\bar q} \ell$. The net effect
of the $U_1$ LQ depends on the value of $\lambda_{U_1 S}$. For
$\lambda_{U_1 S} = 3$, we find that the minimum value of $M_S$ for
which the DM data can be explained is reduced from $\simeq 1.6$ TeV to
$\simeq 1.3$ TeV. As was the case with heavy DM, future indirect
detection measurements will be able to place stringent constraints on
this model.

\bigskip
{\bf Acknowledgments}: We thank Diego Guadagnoli and M\'eril Reboud
for discussions about related topics. This work was partially funded
by the program ``Aide aux \'etudiants et d\'eveloppement international
de la R\'egion Auvergne Rhone-Alpes,'' project number
1900811602-40889, by RFBR and CNRS, project number 20-52-15005 (GB,
AP), and by NSERC of Canada (JK, DL). JK is financially supported by a
postdoctoral research fellowship of the Alexander von Humboldt
Foundation.

\end{document}